# Electric field-induced colossal electroresistance and its relaxation in multiferroic $La_2NiMnO_6$

A. Rebello and R. Mahendiran[*]

Department of Physics and NUS Nanoscience & Nanotechnology Initiative

(NUSNNI), Faculty of Science, National University of Singapore,

2 Science Drive 3, Singapore 117542, Singapore

**Abstract**

In this work, we report direct as well as pulsed electric field-induced resistivity switching and its relaxation in a multiferroic insulator $La_2NiMnO_6$. At a fixed base temperature ($T_b$), the dc resistivity switches abruptly from a high to a low value, which is manifested as an upward jump in the dc current density ($J$) when the electric field ($E$) exceeds a threshold value $E_{th}$. The fractional change in the resistance is as much as 70 % at room temperature for $E_{th}$ = 95 V/cm. The $E_{th}$ increases with lowering $T_b$ and follows the relation $E_{th}(T_b) = E_{th}(0)\exp[-T_b/T_0]$, as similar to the behavior found in charge density wave systems. It is shown that the abrupt jump in $J$ vanishes under pulsed electric fields if the period between pulses is long enough. Surprisingly, a step-like increase in $J$ also occurs at a fixed dc electric field ($E_c$) and $T = T_b$, above a threshold waiting time ($t_{th}$). The $t_{th}$ decreases with increasing $E_c$ and $T_b$. Simultaneous measurement of surface temperature during the $J$-$E$ sweep and temporal studies suggest that conductive channels are created in an insulating matrix due to the local self heating, and the coalescence of these channels above a threshold $E$- field or time causes the observed anomalies in $J$. However, the dissipated Joule power ($P = I_{th}^2 R$) at the transition from high to low resistive state in the sample decreases with lowering temperature, which suggests that the Joule heating is the consequence of transition from the high to low resistance state rather than itself a driving force of the non linear electrical transport. In addition, non linear *J-E* characteristics is also found even with a pulsed voltage sweep, which suggests that intrinsic mechanisms other than self heating is still active in this material.

---

[*] Corresponding author – phyrm@nus.edu.sg

## I. INTRODUCTION

Although perovskite manganese oxides (manganites), with the general formula $R_{1-x}A_x MnO_3$ ($R = La^{3+}$, $Pr^{3+}$ etc., and $A = Ca^{2+}$, $Sr^{2+}$ etc.) have become a hot topic of research due to colossal negative magnetoresistance effect discovered in them in early 90's,[1] it is now clear that the resistivities of these materials are sensitive not only to the external magnetic field but also to other external stimuli such as pressure, X-ray radiation, and electric-field ($E$).[2] In particular, manganites which show coexistence of insulating charge-ordered (CO) and metallic ferromagnetic (FM) states show a dramatic variation in the resistivity by the application of different external stimuli. Among them, the discovery of electroresistance effect in $Pr_{0.7}Ca_{0.3}MnO_3$ by Asamitsu et al.[3] deserves a particular attention. It was found that the resistivity decreased abruptly by several orders of magnitude ($\rho(0)/\rho(V) \approx 10^2 - 10^8$), when the applied bias voltage was increased above a threshold voltage, $V_{th}$ (= 100-700 V) at temperatures below $T_N$ = 120 K. This $E$-field-induced insulator-metal transition was observed to be reversible and first order, as suggested by the hysteresis behavior. Because of the large magnitude of the observed effect, it was aptly named as "colossal electroresistance (CER)" and was attributed to the melting of the charge-ordered state rather than to a conventional dielectric break down. Later, Guha et al.[4] found a strong non-linear voltage-current characteristic in the dc current sweep (-20 mA $\leq I \leq$ 20 mA). They have noted that the voltage increases linearly with current initially, but decreases above a threshold current ($I_{th}$), and this behavior is widely known as the negative differential resistance (NDR) effect. The observed non-linear electrical transport was suggested to arise from the formation of filamentary conductive channels in the charge ordered matrix. The CER effect in $Pr_{0.67}Ca_{0.33}MnO_3$ was also found to be accompanied by an enhancement in the magnetization and it was suggested due to the current-induced transformation of charge ordered antiferromagnetic domains into metallic ferromagnetic domains rather than due to filamentary conduction.[5] Very recently, sliding of charge density waves was suggested to be the origin of the CER effect in a similar charge-ordered compound $La_{0.5}Ca_{0.5}MnO_3$.[6] Resistivity switching between a high value ("OFF" state) and a low value ("ON" state) can also be induced by varying the amplitude and polarity of the applied voltage/current pulses, as demonstrated in $Pr_{0.7}Ca_{0.3}MnO_3$ and many other oxides.[7,8,9,10] These observations have generated a lot of interest both in the view point of fundamental physics as well as the possible application of resistivity switching in non volatile resistive random access memory (RRAM).

Besides in manganites, resistivity switching has been reported in a wide range of oxides such as $NiO$,[11] $CoO$,[12] $TiO_2$,[13] Cr-doped $SrTiO_3$,[14] $LuFe_2O_4$,[15] $Na_{0.5-\delta}CoO_2$,[16] $La_{2-x}Sr_xNiO_4$, $La_{0.33}Sr_{0.67}FeO_3$,[17] and nano- ionic materials.[18] The observation of resistivity switching effect in such a wide range of materials pose a considerable challenge to find a single common mechanism. It is possible that different mechanisms are operative in these materials, and depends on the details of the nature of electronics transport of the particular system under consideration. Recent reviews summarize advancements in this field.[19] For manganites alone, several mechanisms have been proposed which include electric-field induced switching of orbital states,[20] trap controlled space-charge limited current,[21] excitation of charge density waves,[22] small-polaron-hopping,[23] phonon-assisted electron delocalization[24], doping control of electric state at the interface,[25] Mott transition at the interface,[26] pulse-generated crystalline defects and migration of oxygen ions.[27] A recent phenomenological approach involves a nonpercolating domain structure as the origin of the resistivity switching.[28] So, a coherent picture on the origin of the CER effect in manganites is yet to emerge.

There has always been some concern about the role of resistive Joule heating in the observed CER effect since significant Joule heating can lead to a decrease in resistance if the sample is semiconducting ($d\rho/dT < 0$). The currents necessary for non-linear effects in manganites are generally in the order of $I$ = 20-100 mA and in this regard, distinguishing intrinsic mechanisms of the CER effect from those originating from thermal effects remains intriguing and challenging. While majority of the researchers have assumed that increase in the temperature of the sample is rather small (< 5 K) during the current sweeps, a few researchers reported significant increase of temperature ($\Delta T \approx$ 30-250 K), by direct measurement using a temperature sensor attached close to the sample or on the sample itself [29,30,31] or estimated from simple models of heat conduction.[32] Tokunaga *et al.*[33] demonstrated that the application of a large current (~30 mA) in phase separated $(La_{0.3}Pr_{0.7})_{0.7}Ca_{0.3}MnO_3$ leads to the collapse of percolative conduction paths through local Joule heating of metallic channels which in turn results in a positive electroresistance (resistance increase under an electric field) instead of the usual negative electroresistance effect. They have also reported unusual current oscillation phenomena under a constant dc voltage bias and thermally driven non volatile resistive memory effect.[34] However, it is premature to conclude that the CER effect is caused by the Joule heating alone. Instead, extensive and unbiased investigations are necessary to elucidate mechanisms of the CER effect in any new material. It is mandatory, rather than optional, to monitor the temperature of the sample itself during the current-voltage sweep instead of taking note of the values recorded by the temperature sensor located away from the sample as in commercial cryostat such as the Physical Property Measuring System (PPMS). In addition, it is important to investigate non-linear conduction with pulsed current/voltages instead of the normally used dc current/voltage sweep.[35]

The ferromagnetic insulator $La_2NiMnO_6$ (LNMO) has recently attracted more attention due to the discovery of a large change in the dielectric constant under a magnetic field ("magneto-dielectric effect").[36] Such multiferroic materials are promising for applications in the next generation spintronic devices for data storage and new types of magnetic sensors. If electroresistance can also exist in this compound, it can add to its versatile and multifunctional capabilities. However, nonlinear electric transport in this compound has not been reported so far and this is the motivation of the present work. In this work, we report the current density versus electric field (*J-E*) characteristics in the ferromagnetic insulator, $La_2NiMnO_6$ (LNMO) both in dc and pulsed voltage sweep mode. In addition to the nonlinear electrical transport, we also study the magnetocaloric effect in this compound.

## II. EXPERIMENTAL DETAILS

The $La_2NiMnO_6$ (LNMO) sample was prepared by a soft chemical route. The samples were synthesized by employing citrate-gel method where, aqueous solutions of $La(NO_3)_3.9H_2O$ (0.011 mol, Merck, 99.9%), $Mn(NO_3)_2.4H_2O$ (5.5 mmol, Merck, 99.9%) and $Ni(NO_3)_2.6H_2O$ (5.5 mmol, Merck, 99.9%) were prepared in required stoichiometry, and mixed with citric acid solution (aqueous) in twice the molar ratio of the total metal ions. The solution was then refluxed for four hours. The solvent was slowly evaporated off from this well-mixed solution, until a gel is formed. This gel was then decomposed directly at 623 K. The bluffy mass produced was then ground and annealed at 873 K for 12 hours in order to remove the carbon left out in the as decomposed sample. Further annealing was done at 1073 K for 12 hours, 1273 K for 24 hours and finally a pellet was fired at 1473 K for 24 hours, with intermediate grinding at every step. The X-ray diffraction characterization was performed using Cu Kα radiation on a Philips Xpert powder diffractometer and the magnetic transition temperature was determined using a Vibrating Sample Magnetometer (Quantum Design Inc, USA).

The four probe *dc* and pulsed *J-E* characteristics of a bar shaped polycrystalline LNMO sample (length = 10 mm, width = 3 mm and thickness = 2.5 mm) were measured using source-measure units (Keithley 2400 and Yokogawa GS610) interfaced to the Physical Property Measurement System (PPMS, Quantum Design Inc, USA). The electrical contacts were made with Ag-In alloy or Ag paint and the results were found to be identical. While sweeping the voltage up to ± 80 V, the current compliance was set to ±20 mA in order to protect the sample from any electrical damage. The sample was glued to a thin mica substrate, with GE-7031 varnish, which was attached to the standard Au plated copper sample puck. To monitor the surface temperature ($T_S$) of the sample, a Pt-100 thermometer of size 3x2x1.2 mm$^3$ was thermally anchored to the top surface of the sample with good thermal conductive grease (Apiezon-N grease) and a small quantity of GE-7031 varnish. The Pt resistor was positioned between two voltage probes that were separated by 5.5 mm. The four probe resistance of the Pt thermometer was monitored by measuring the voltage across the Pt-resistor using a Keithley 2182A nanovoltmeter while supplying a constant current of 10 µA with Keithley 6221 dc and ac current source. The cernox sensor attached to the sample puck through the thermal sink inside PPMS measured the base temperature ($T_b$). It is to be pointed out that the base temperature recorded by the PPMS did not show any variation (remains at the stable value $T_b$) with $E$-field sweep and this limits the measurement of actual response of the surface temperature of the sample. On the contrary, since the top surface of the sample is in very good thermal contact with the Pt100 thermometer in our experimental set up, actual response of the surface temperature of the sample during the voltage sweeps can be measured in our experimental set up. Although a small temperature difference between thermometer and the sample is possible, the difference will be one tenth of Kelvin and hence negligible. All the pulsed measurements were performed using a Yokogawa GS610 source-measure unit.

### III. RESULTS AND DISCUSSION

#### A. Magnetization and magnetocaloric effect

Figure 1(a) shows the temperature dependence of the magnetization, $M(T)$, of the LNMO sample measured at $H = 1$ kOe. The $M$-$T$ data suggests that the sample undergoes a paramagnetic to a ferromagnetic transition upon cooling and the Curie temperature ($T_C$) identified from the inflection point of the $M$-$T$ data is $T_C = 275$ K. The $M$-$H$ curve at $T = 10$ K, shown in the inset of Fig. 1(a) indicates a typical ferromagnetic hysteresis loop behavior with a small coercivity ($H_C =$ 200 Oe). The saturation magnetization ($M_s$) obtained from the extrapolation of the high field magnetization to the origin ($H = 0$ Oe) is $M_s = 4.2$ $\mu_B$ which is lower than the theoretical value of 5 $\mu_B$ assuming a complete ordering of $Ni^{2+}$ (S = 1) and $Mn^{4+}$ (S = 3/2) spins. The isothermal $M$ vs. $H$ curves at close temperature interval of 5 K between 350 and 220 K are shown in Fig. 1 (b). Although the $T_C$ determined from the $M$-$T$ curve is 275 K, the M-H curve shows a linear behavior only above 330K. Between $T_C$ and 330K, the low-field magnetization is highly non linear with $H$, which indicates a super paramagnetic like behavior. It is possible that superparamagnetic clusters coexist with majority paramagnetic phase for certain temperature range above $T_C$. From the isothermal magnetization curves recorded at close temperature intervals, we have calculated the magnetic entropy change (-$\Delta S_m$) and plotted it in Fig. 1(c) for different values of the magnetic field. The –$\Delta S_m$ shows a peak around $T_C$ and the magnitude of the peak increases with $H$. A maximum value of $\Delta S_m = -1.7$ J/Kg K around $T_C$ is found for $\Delta H = 5$ kOe. The observed value of the $\Delta S_m$ in $La_2NiMnO_6$ is smaller than the values found in other ferromagnetic CMR manganites.[37] If we assume random alignments of $Ni^{2+}$, and $Mn^{4+}$ spins in the paramagnetic state, the expected spin entropy change is, $\Delta S = Nk_B \ln(2J+1) \approx 32$ J/kgK. However, the estimated entropy change from the experimental magnetization data is much smaller and suggests that magnetic entropy can be increased at higher magnetic fields. It is not clear why the magnetic entropy in this compound is lower than in other CMR manganites and therefore is a topic which has to be investigated in detail. However, we leave this topic here and focus more on the nonlinear electrical transport in this paper.

### B. Nonlinear electrical transport with dc current

Figure 2(a) shows temperature dependence of the resistivity ($\rho$) of LNMO sample at four different current strengths, $I = 10$ µA, 100 µA, 1 mA and 10 mA. When $I = 10$ µA, the sample shows a semiconducting behavior while decreasing $T$ from 400 K. The X- axis shows the temperature ($T_b$) of the sample recorded by the PPMS, which we call as the "base temperature". The high temperature ($T > 225$ K) resistivity curve for $I = 10$ µA is fitted to the Arrhenius model ($\rho = \rho_0 \exp[E_A/(k_B T_b)]$), as shown in the inset of Fig. 1(a). It suggests thermally activated electrical transport in the sample with an activation energy of $E_A = 0.27$ eV. The resistivity for higher current strengths also initially increases with lowering temperature from 400 K, but they differ very much below 300 K. The resistivity for $I = 10$ mA and at 200 K is two orders of magnitude smaller than the value at 1 mA at the same temperature. Fig. 2(b) shows the sample surface temperature ($T_s$) measured by the Pt resistor attached to the sample surface. While the $T_s$ for $I = 10$ µA closely matches with $T_b$ between 400 and 225 K, the $T_s$ for $I = 10$ mA does not match with $T_b$ even at 400 K and shows a significant difference at low temperatures. The current-induced electroresistance, $ER$ (%) = $[\rho(100$ µA$) - \rho(10$ mA$)]/\rho(100$ µA$) \times 100$ increases from 49 % at 400 K to 98 % at 225 K as shown on the right scale of Fig. 2(b). Although similar electroresistance effect was reported earlier in $Nd_{0.5}Ca_{0.5}MnO_3$, which was attributed to the current-induced depinning of randomly pinned charge solid, the response of the sample surface temperature was not explicitly measured.[38]

In Fig. 3, we show the current density ($J$) versus electric field ($E$) behavior of the LNMO sample at different temperatures measured in voltage sweep mode. At $T_b = 350$ K, $J$ varies gradually with $E$ below 20 V/cm, but increases steeply above a threshold field, $E_{th} = 35$ V/cm. Upon reducing the $E$ from the maximum value, $J$ shows hysteresis at higher fields but the curves merge near the origin. The $J$-$E$ curves at higher temperatures also show strong nonlinearity and much broader hysteresis. The threshold $E$-field increases from $E_{th} = 35$ V/cm at $T_b = 350$ K to almost $E_{th} = 150$ V/cm at $T_b = 280$ K. In Fig. 3(b), we show the change in $T_s$ during the electric-field sweep. At $T = 350$ K, the surface temperature increases from $T_s = 350$ K at $E = 0$ V/cm to $T_s = 380$ K at $E = 60$ V/cm. The rapid increase in $T_s$ is accompanied by the rapid increase in the current density. The $T_s$ shows a concomitant change with the current density during the $E$-field sweep at low temperatures too. The lower the $T_b$, the larger is the $E_{th}$ and also is the $T_s$. When the $E$-field is reduced to zero, the temperature remains nearly constant down to certain $E$-field and then decreases resulting in a huge hysteresis similar to the current density. Although the temperature of the sample, after the $E$- field is returned to zero, is appreciably higher than the starting value, the difference in the current density around the origin is small and is not visible in the scale of the graph.

We plot the threshold field ($E_{th}$) obtained from Fig. 3(a) as a function of temperature in the main panel of Fig. 4(a). The $E_{th}$ increases exponentially with decreasing temperature. The line is guide to eye. The plot of $\ln(E_{th})$ versus temperature is linear as shown in the inset of figure. We estimate the power ($P = I_{th}^2 R$) needed for the abrupt increase of the current density. The power needed to cause high- to low- resistivity transition decreases nearly exponentially with temperature as shown in Fig. 4(b).

We have also investigated the effect of interrupting the $E$- field sweep on the current density and temperature. We show the effect of partial $J$-$E$ loops obtained by sweeping the dc $E$– field up to a maximum field $E_{max}$ at 290 K in Fig. 5. Four different $J$-$E$ curves were taken with $E_{max}$ = (a) 19 , (b) 57, (c) 76 and (d) 152 V/cm, where $E_{max}$ is the maximum field in each sweep (0 →$E_{max}$→0 V/cm). Figure 5 shows the *dc J-E* curves on the left panel and the response of the surface temperature on the right panel, where the forward (0→ $E_{max}$ V/cm) and backward ($E_{max}$ →0 V/cm) sweeps are shown in the open (black) and closed (red) symbols respectively. When $E_{max}$ = 19 V/cm, the $J$-$E$ curve is linear in both the forward and backward sweep (Fig. 5 (a)) and the $T_s$ shows a negligible change during the sweep (Fig. 5(e)). When $E_{max}$ is increased to 57 V/cm, the *J-E* curve (Fig. 5(b)) shows a small hysteresis and nonlinearity in both the up and down sweep at high $E$ fields. The corresponding surface temperature of the sample increases from $T_s$ = 290 K at 0 V/cm to $T_s$ =300 K at 57 V/cm as can be seen in Fig. 5(f). The *J-E* curve shows an anomalous behavior when the $E_{max}$ = 76 V/cm, as shown in Fig. 5(c). The current density shows a gradual increase from 0 to 50 mA/cm$^2$ during the forward sweep of $E$-field from 0-76 V/cm. However, when the $E$ is reduced from $E_{max}$, $J$ shows an unexpected dramatic increase rather than a decrease. After the abrupt increase, $J$ remains nearly constant down to $E = 25$ V/cm and then decreases to zero as $E \rightarrow 0$ V/cm. The dramatic changes in $J$ are accompanied by similar changes in the $T_s$, as can be seen in Fig. 5(g). In the final sweep with $E_{max} = 152$ V/cm, the current shows an abrupt jump just above $E_{th} = 76$ V/cm in the forward sweep and remains at a high value for $E > E_{th}$. During the downward sweep, $J$ remains nearly constant down to $E = 25$ V/cm, and then drops to zero as $E \rightarrow 0$ V/cm. The corresponding surface temperature also shows similar behavior to the current density with a huge increase from $T_s = 300$ K at $E = 0$ V/cm to $T_s = 380$ K at $E = 95$ V/cm.

## C. Electrical transport with pulsed voltage

Now let us turn our attention to the effect of pulsed electric field. In Fig. 6(a), we compare the $J$-$E$ characteristics at $T_b = 300$ K in the dc (open circle) and pulsed (line) modes. The corresponding response of the $T_s$ is shown on the right scale. For pulsed field sweep, we used pulses of a long period ($P_D = 5$ s) and a short pulse width ($P_W = 0.025$ s). In the *dc E*-field sweep, the current density initially increases gradually, but a rapid increase occurs just above the threshold field, $E_{th} = 95$ V/cm which implies about 71 % decrease in the resistivity. A huge hysteresis results when the $E$- field is reduced to zero. The surface temperature of the sample also shows a hysteretic behavior and an increase from $T_s = 300$ K at $E = 0$ V/cm to $T_s = 380$ K at $E = 95$ V/cm. On the contrary, the $J$ increases gradually with $E$ in the pulsed mode, the increase in $J$ being small compared to the dc mode. The surface temperature of the sample shows a negligible change of one tenth of K in the pulsed mode. However, the pulsed $J$-$E$ curve shows a change of slope at $E = 45$ V/cm and this nonlinearity results in ≈ 22 % decrease in the resistivity when field is changed from $E = 0$ V/cm to $E = 95$ V/cm. Figure 4(b) (left scale) shows the comparison between the dc and pulsed $J$-$E$ curves below room temperature, $T_b = 250$ K. The dc $J$-$E$ curve at $T = 250$ K does not show a strong non linear behavior unlike at 300 K because the resistance of the sample has increased and applied voltage is small enough to induce the low resistive state. Nevertheless, it is worthy to note that the pulsed *J-E* curve at $T = 250$ K is similar to the pulsed *J-E* curve at $T = 300$ K (Fig. 4(a)), with a change of slope and nonlinearity. A noticeable change in $T_s$ occurs in the dc mode, but is negligible in the pulsed mode as shown on the right scale of Fig. 4(b).

In Fig. 7(a), we compare the pulsed *J-E* curves measured at different base temperatures and the concomitant changes in the surface temperature are shown in Fig. 7(b). At all base temperatures, the *J-E* curves exhibit a clear change in the slope. When $T_b = 375$ K, the pulsed *E*-field increases the surface temperature of the sample and hence the non-linearity in *J* is more pronounced. This is because of the lower resistance of the sample at higher temperature. Hence, at the maximum *E* field, the current is larger and the power dissipation ($I^2R$) increases.

## D. Time dependence of the dc resistivity

The abrupt increase in the current during the *E*-field down sweep (76-0 V/cm), as observed above (Fig. 5(c)), has directed us to study current relaxation behavior in the sample by measuring the current density as a function of time for a constant *dc* field at $T_b = 300$ K. We show the temporal evolution of the current density for a few fixed fields, $E_c = 58, 59, 60, 61, 63, 65, 67,$ and 75 V in Fig. 8(a). When $E_c = 58$ V/cm, the current density shows a gradual increase in the beginning but remains at a low value ($J = 30$ mA/cm$^2$) for a long time of 6000 s. We have shown the data only up to 2000 s in the figure for easy comparison with other data. However, when $E_C = 59$ V/cm, after a small initial growth *J* remains nearly a constant below 900 s, but a step like increase occurs at a threshold time, $t_{th} = 1000$ s. As the magnitude of the $E_c$ increases, the abrupt jump in *J* shifts down to smaller time. The abrupt jump in *J* is invariably accompanied by a similar trend in $T_s$ as shown in Fig. 6(b). For instance, $T_s$ for $E_c = 59$ V/cm shows a gradual increase below $t = 900$ s, then increases sharply from $T_s = 310$ K at $t = 900$ s to $T_s = 390$ K at $t =$ 1000 s. The sharp increase in $J$ and $T_s$ occurs at a lower threshold time, $t_{th}$ = 800 s for $E_c$ = 61 V/cm and at higher field values, $t_{th}$ again decreases.

Figure 9 compares the *dc* resistive relaxation at a temperature below ($T_b$ = 280 K) and above ($T_b$ = 350 K) the room temperature. At $T_b$ = 280 K (Fig. 7(a)), the current density shows an initial gradual increase and then remains at a low value up to 2000 s for $E_c \leq$ 110 V/cm. *It is surprising that J shows an abrupt jump after $t_{th}$ = 1100 s for an increase of just 2 V/cm, i.e. at $E_C$ = 112 V/cm.* The data also shows that higher the $E_c$ smaller the $t_{th}$. A similar temporal evolution of $J$ is also found at 350 K (Fig. 9 (b)). However, the $E$-field needed to induce the low resistive state is smaller than at 280K and it also occurs at a smaller $t_{th}$. A comparison of Figs. 8 and 9 clearly indicate that $E_{th}$ and $t_{th}$ increases with lowering temperature. The corresponding changes in the $T_s$ during the relaxation measurements at $T_b$ = 280 and 350 K are shown in Figs. 7(c) and 7(d) respectively which suggest that the transition to the low resistive state is accompanied by a rapid increase in the surface temperature of the sample. The insets show the energy, $W_J = \int_0^{t_{th}} VI dt$ involved in the transition, which varies nearly linearly with the threshold time.

The above results are reproducible at ambient environment outside the cryostat as shown in Fig. 10. Here too, we find that an abrupt change in $J$ is induced by a small increment in the $E$-field as can be clearly seen in the behavior at $E_c$ = 48 V/cm and 50 V/cm. The $t_{th}$ is also very sensitive to a small change in $E_c$. The increase in temperature which accompanies the resistive switching is plotted in Fig. 10 (b) and the inset shows the variation of $W_J$ with $t_{th}$.

## IV Discussion

Besides in the charge ordered manganites, an abrupt increase in the current density above a threshold electric field was also reported in charge density wave (CDW) system such as NbSe$_3$, K$_{0.3}$MoO$_3$[39], La$_{2-x}$Sr$_x$NiO$_4$,[40] and organic conductor MDT-TS[41] in which the relation $E_{th}(T) = E_{th}(0)\exp[-T_b/T_0]$ holds true. The linear behavior of $\ln(E_{th})$ vs. $T_b$ curve (discussed earlier in Fig. 3(a)) is interpreted in the literature due to the depinning of charge density waves by the applied electric field. Recently, evidence in favor for the electric-field-induced sliding of charge density waves in $La_{0.5}Ca_{0.5}MnO_3$ was presented.[6] An abrupt resistivity switching similar to the titled compound was also recently found in $LuFe_2O_4$ and was attributed to electric field-driven collapse of charge ordering.[42] While the electric field can destroy charge ordering and can cause collective transport of otherwise immobilized electrons in these system, contribution from Joule heating should not be ignored especially when high current density is involved.

Our combined measurements of the $J$-$E$ characteristics and $T_s$ clearly indicate that there is significant self-heating in the vicinity of $E_{th}$ in $La_2NiMnO_6$. A simple scenario is to consider that local Joule heating initially causes formation of low resistive filaments in the otherwise insulating matrix. Once the $E$-field exceeds $E_{th}$, each filament reconfigures themselves in to stripes and bridges the gap between the electrodes as in a dielectrophoresis system.[43] This leads to a percolation of conducting channels and triggers a surge in the current. The heat balance equation is $C\frac{dT_s}{dt} = -K(T_s - T_b) + W$ where $C$ is the heat capacity of the sample, $K$ is the effective thermal conductance of the sample to the thermal bath, $T_b$ is the temperature of the base, $T_s$ is the surface temperature of the sample ($T_s = T_b$, if there is no Joule heating) and $W = I^2R$ is the electric power which is converted into Joule heat. In the steady state $dT/dt = 0$, and hence the temperature of the sample is $T_s = T_b + \frac{R(T)I^2}{K}$. If the dissipated power in the sample is sufficient to warm up the sample and its immediate surrounding, temperature of the sample will rapidly rise from the base temperature $T_b$. Since the sample is semiconducting, the global heating of the sample above the threshold field $E_{th}$, leads to a sudden decrease in the resistance and increase in $J$ as observed.

The dramatic increase in the current density and $T_s$ after a threshold time $t_{th}$, reflects competition between the power supplied to the sample, heat dissipated in the sample and heat transferred to the surrounding. Tokunaga *et al.*[33] has observed an abrupt increase in the resistivity (opposite to the behavior in our sample) after a threshold time for different current strengths at T = 30 K in the Cr- doped $Nd_{0.5}Ca_{0.5}MnO_3$, where the transition is governed by the energy applied to the sample rather than by the magnitude of the current. Following their approach, we have estimated the net energy needed to cause the transition at $T_b$ = 300 K in our sample as $W_{th} = \int_0^{t_{th}} [VI - K(T_s - T_b)]dt$, where $K$ is the thermal conductance to a thermal bath, and the energy of joule heating $W_J = \int_0^{t_{th}} VIdt$ is calculated from the experimental data. The $W_J$ versus $t_{th}$ curve shows a linear behavior as shown in the inset of Fig. 6(a), where we approximate the cooling term $K(T_s - T_b)$ by a constant value $p_c$, i.e., $W_J = W_{th} + p_c t_{th}$. From the intercept and the slope of $W_J$ versus $t_{th}$ curve, we obtain $W_{th}$ = 0.8 J and $p_c$ = 80 mW. The $W_J$ versus $t_{th}$ curves at other temperatures also show a linear behavior as shown in the insets of Figs. 6(c) and 6(d). It has to be noted that the transition is caused by a *E*-field value larger than a minimum field bias, i.e. $E_c \geq$ 58 V/cm at *T* = 300 K, similar to the observation at *T* = 30 K in the phase separated Cr-doped $Nd_{0.5}Ca_{0.5}MnO_3$ sample by Tokunaga *et al.*[33] The sharp jumps in the current density in our sample can be interpreted as due to nucleation of such localized conductive filaments. The propagation of heat through the sample will cause more filaments to nucleate which will grow in time and coalesce at $t_{th}$, thereby resulting in the sharp increase in the current density as observed.

However, there are certain aspects which indicate that Joule heating alone is not responsible for the non linear electrical transport in our sample. First, we find that the non-linear *J-E* behavior is found even with pulsed electric fields even though Joule heating is negligible in this case. Second, an abrupt jump in the current as a function of waiting time occurs only for electric fields, $E \geq E_c$. The $t_{th}$ is also very sensitive to small changes in $E_c$. Third, the power dissipated in the sample ($P= I_{th}^2 R$) to initiate the abrupt increase in $J$ decreases with lowering temperature. These observations lead to the following question: Is the Joule heating driving force behind the abrupt increase $J$ or is it the consequence of the transition from a high to low resistive state where the increase in $J$ is triggered by some intrinsic mechanism? Recently, nonlinear electron transport in insulating state of $La_{0.88}Ca_{0.18}MnO_3$ was ascribed due to "hot electron" effect where the high input electrical power decouples the electron and lattice baths which leads to rapid heating of the electron bath above the phonon bath.[44] The electrical conduction is dominated by "hot electrons" having higher average kinetic energy than the rest of electrons. The increase in the temperature of the sample can't be easily detected in short time scale unless hot electrons transform their thermal energy to the phonon bath. Based on the above scenario, it is also possible that the resistivity switching from high to low resistance state is initiated by hot electron transport below $E_{th}$ in our sample. Once $E$- field increases above $E_{th}$, the inelastic collisions of electrons with phonon leads to a rapid increase of the sample's temperature and hence decreases the resistivity.

## IV. CONCLUSIONS

In summary, we have studied pulsed as well as *dc* voltage induced electroresistance in the multiferroic ferromagnetic semiconductor $La_2MnNiO_6$. It is found that current density increases abruptly above a threshold dc electric field $E_{th}$ at a fixed base temperature. The $E_{th}$ increases exponentially with lowering temperature and can be described by an equation similar to the one used for the charge density wave system. Anomalous hysteresis in the *J-E* characteristics is found if the dc electric field is partially reversed from a maximum value. The abrupt increase in the current density vanishes in the pulsed electric field sweep. Surprisingly, an abrupt increase in the current was also found at a fixed base temperature after a certain waiting time. Simultaneous measurement of the sample surface temperature during the *J-E* sweep suggests that the abrupt increase in the current is accompanied by a sudden increase in the sample temperature. Although our results can be interpreted as due to the formation of low-resistive filaments in an otherwise insulating matrix and percolation effect, there are several other aspects, particularly the sensitivity of $t_{th}$ to $E_c$ and behavior of *J-E* with pulsed electric field, which can not be solely explained on the basis of Joule heating alone. We believe that other experimental technique such as dynamical electrical transport (ac electric field superimposed on a linearly increasing dc electric field) [6] may shed light on the non linear electrical transport in our compound.

**Acknowledgements:** R. M thanks the Ministry of Education (Singapore) and Deputy President Office of Research & Technology, NUS for supporting this work through the grants ARF-Tier1-R144-000-167-112 and R144-000-197-123.

**Figure captions:**

**Fig. 1**: (Color online) (a) Temperature (*T*) dependence of the magnetization (*M*) measured with $\mu_0H$ = 0.1 T. The inset shows the *M* vs. *H* curve at *T* = 10 K. (b) shows the magnetization isotherms at each 5 K interval in the temperature range 350-220 K. (c) shows the magnetic entropy change (-$\Delta S_m$) estimated from the magnetization isotherm curves.

**Fig. 2**: (Color online) Temperature dependence of the: (a) resistivity ($\rho$) at different current strengths *I* = 10 µA, 100 µA, 1 mA, and 10 mA. The inset shows the Arrhenius fit at *I* = 10 µA. (b) surface temperature, $T_s$ on the left scale and the percentage electroresistance (ER) for *I* = 10 µA and 10 mA on the right scale.

**Fig. 3**: (Color online) (a) Current density (*J*) *vs*. electric field (*E*) curves at different temperatures and (b) the respective response of the $T_s$ during the field sweep. The voltage sweep rate was 0.02V/sec in all the measurements.

**Fig.4:** (a) (Color online) Temperature dependence of the threshold electric field $E_{th}$ in the resistivity switching, extracted from figure 3(a). The inset shows the fit to the equation $E_{th}(T) = E_{th}(0)\exp[-T/T_0]$. (b) shows the temperature dependence of the input power ($P = I_{th}^2R$) to cause the resistivity switching.

**Fig. 5**: (Color online) The *dc J-E* curves (0→$E_{max}$→0 V/cm) successively taken with $E_{max}$ = 19, 57, 76 and 152 V/cm at 300 K. The arrows indicate the direction of the sweep.

**Fig. 6**: (Color online) The *dc* and pulsed current density (*J*) dependence of the electric field (*E*) and the corresponding change in the $T_s$ during the field sweep at (a) *T* = 300 K and (b) *T* = 250 K.

**Fig. 7**: (Color online) (a) The *J-E* isotherms measured with pulsed electric field at different base temperatures and (b) respective change in the surface temperature, $T_s$, during the pulsed E-field sweep.

**Fig.8 :** (Color online) Temporal dependence of the (a) current density ($J$) and (b) corresponding change in $T_s$ for fixed bias fields, $E_c$ = 58, 59, 60, 61, 63, 65, 67, and 75 V/cm at 300 K. The inset shows the dependence of $W_J$ on the threshold time $t_{th}$.

**Fig. 9**: (Color online) Temporal dependence of the current density for different $E_c$ at $T$ = 280 and 350 K is plotted in (a) and (c) respectively. The corresponding change in $T_s$ is plotted in (b) and (d). The inset shows the respective dependence of $W_J$ on the threshold time $t_{th}$.

**Fig. 10**: (Color online) Temporal dependence of the (a) current density and (b) corresponding change in $T_s$ for fixed bias voltages, $E_c$ = 48, 50, 51, 53, 57, and 76 V/cm at ambient environment at room temperature. The inset shows the dependence of $W_J$ on the threshold time $t_{th}$.

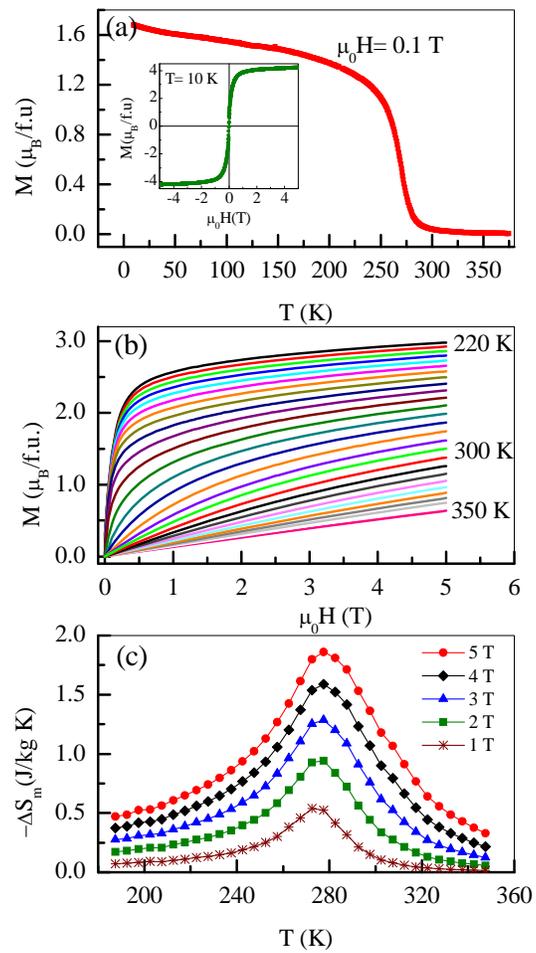

Fig. 1

Rebello *et al*.

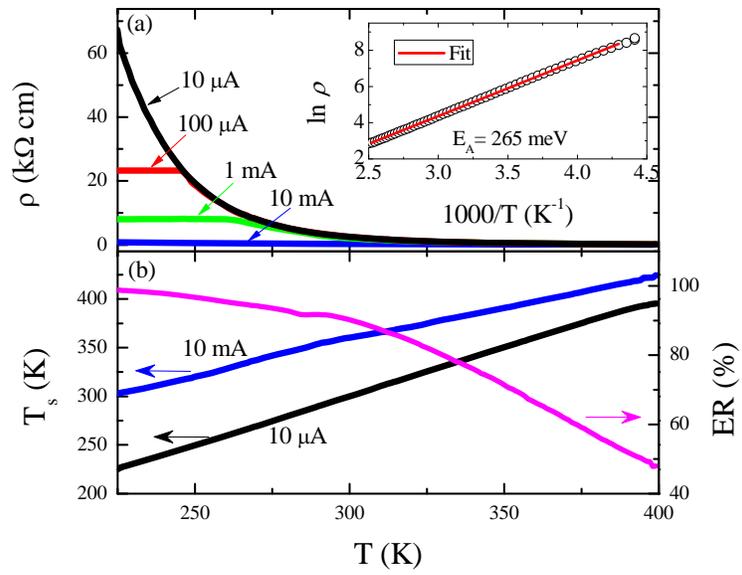

Fig. 2

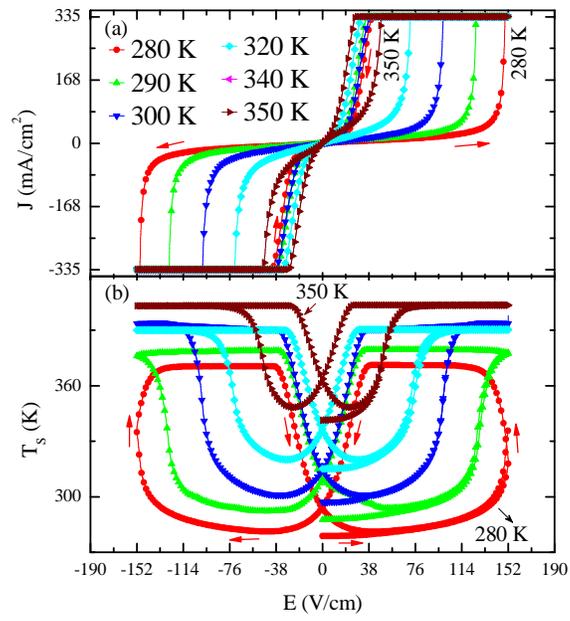

Fig. 3

Rebello *et al*.

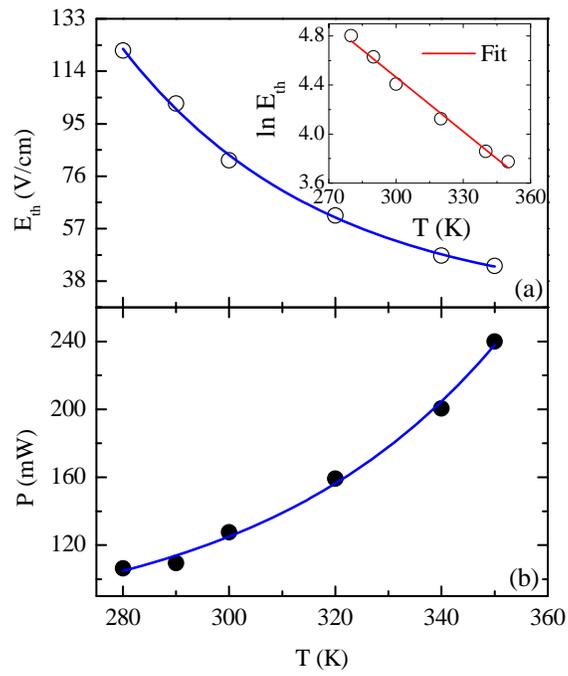

Fig. 4

Rebello *et al*.

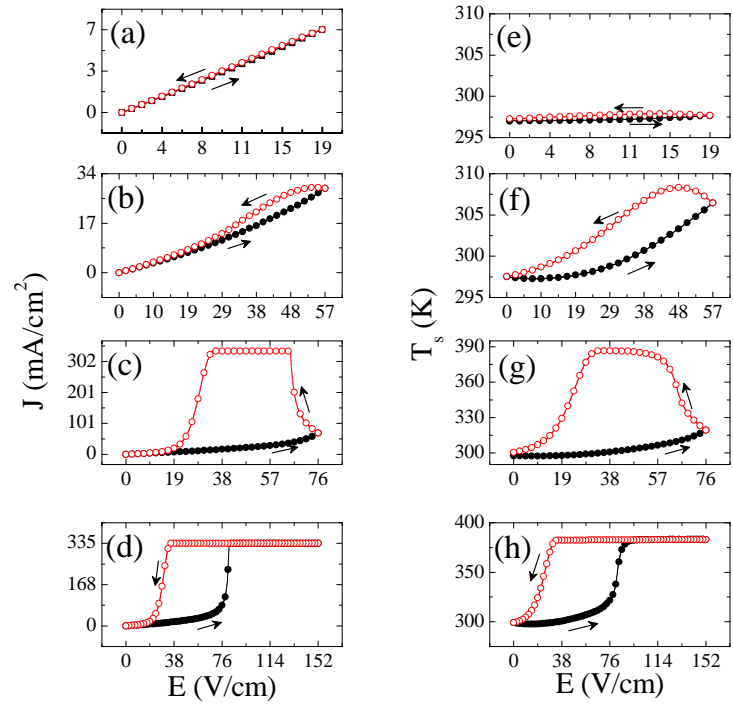

Fig. 5

Rebello *et al*.

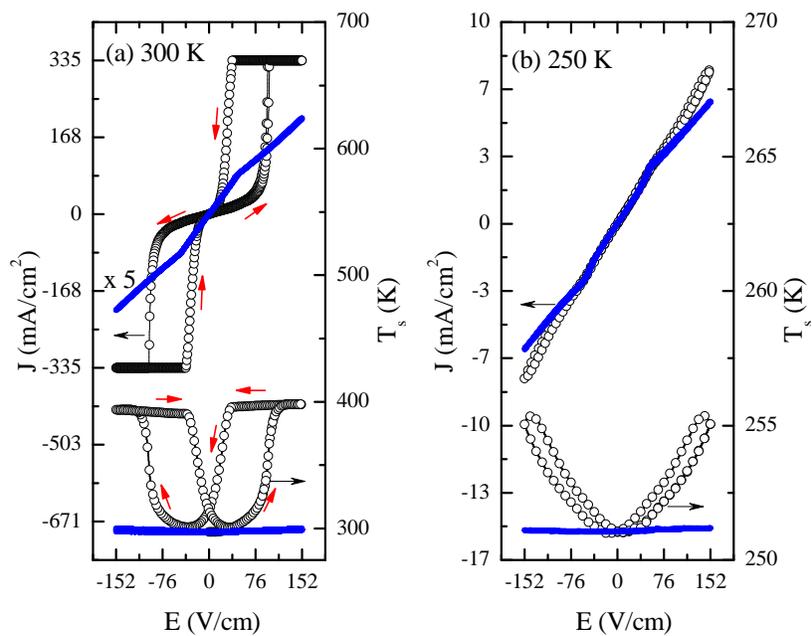

Fig. 6

Rebello *et al*.

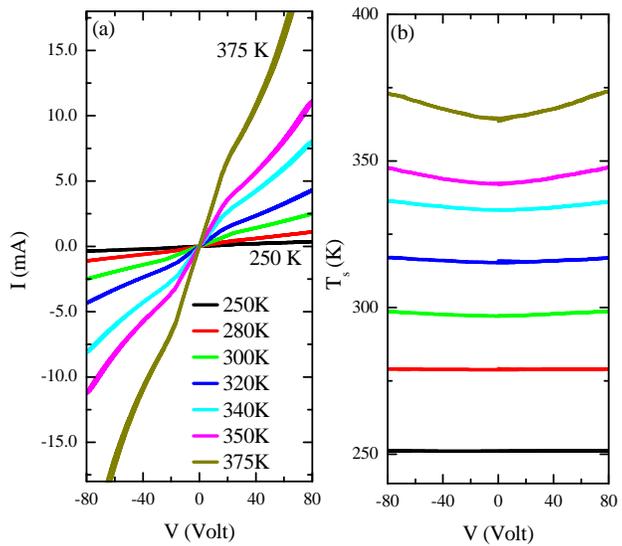

Fig. 7

Rebello *et al*.

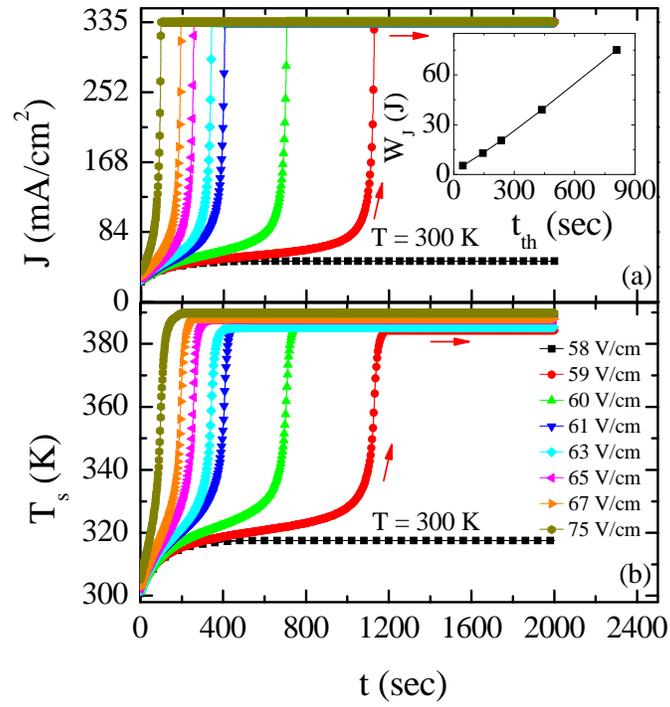

Fig. 8

Rebello *et al*.

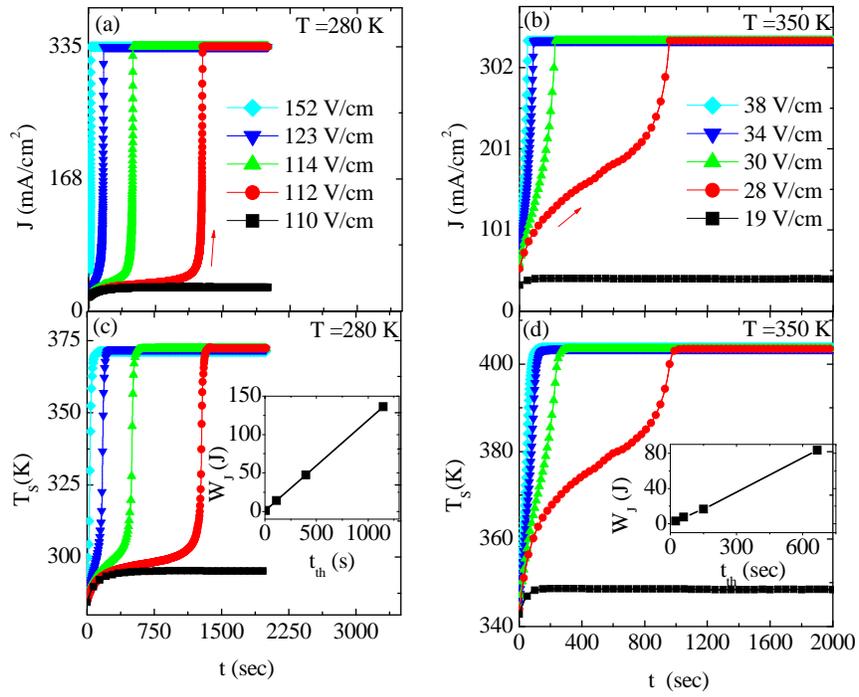

Fig. 9

Rebello *et al*.

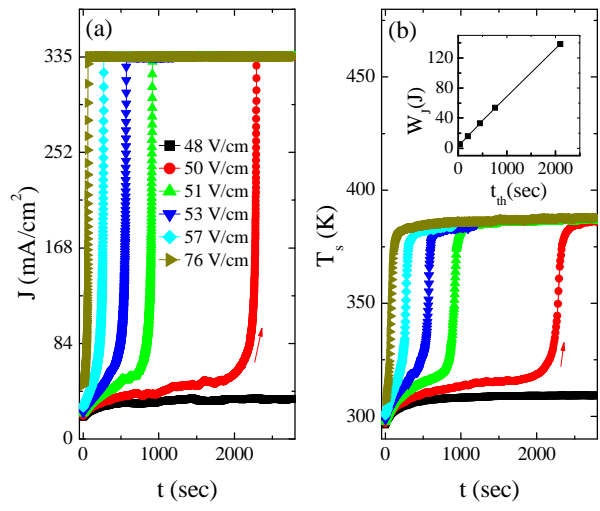

Fig. 10

Rebello *et al*.